\documentclass[useAMS,usenatbib]{mn2e}
\usepackage{amsmath}
\usepackage{hyperref}
\usepackage{caption}
\usepackage[toc,page]{appendix}
\usepackage{pslatex}
\usepackage{amsmath}

\usepackage{graphicx}
\usepackage{float}

\title[Effect of Dust Evaporation and Thermal Instability]{Effect of Dust Evaporation and Thermal Instability
on Temperature Distribution in a Protoplanetary Disk}
\author[Pavlyuchenkov et. al.]{Ya. N. Pavlyuchenkov, V. V. Akimkin, A. P. Topchieva, E. I. Vorobyov  \\
        Institute of Astronomy of the Russian Academy of Sciences, Moscow, 119017 Russia}
\date{}
\begin{document}
\date{Received 16.01.2023; revised 22.03.2023; accepted 27.03.2023 \\
pavyar@inasan.ru}


\maketitle

\begin{abstract}
\noindent
The thermal instability of accretion disks is widely used to explain the
activity of cataclysmic variables, but its development in protoplanetary
disks has been studied in less detail. We present a semi-analytical
stationary model for calculating the midplane temperature of a gas and
dust disk around a young star. The model takes into account gas and dust
opacities, as well as the evaporation of dust at temperatures above
1000~K. Using this model, we calculate the midplane temperature
distributions of the disk under various assumptions about the source of
opacity and the presence of dust. We show that when all considered
processes are taken into account, the heat balance equation in the region
$r<1$~au has multiple temperature solutions. Thus, the conditions for
thermal instability are met in this region. To illustrate the possible
influence of instability on the accretion state in a protoplanetary
disk, we consider a viscous disk model with $\alpha$-parameterization of
turbulent viscosity. We show that in such a model the disk evolution is
non-stationary, with alternating phases of accumulation of matter in the
inner disk and its rapid accretion onto the star, leading to an episodic
accretion pattern. These results indicate that this instability needs to
be taken into account in evolutionary models of protoplanetary disks.
\\
\\
{\bf DOI:10.1134/S1063772923050086}
\\
\\
\end{abstract}

\section{Introduction}

The thermal structure of a protoplanetary disk of gas and dust around a young star is inextricably linked with many key processes that affect the evolution of the disk itself. Temperature determines the physical, chemical and ionization structure of the protoplanetary disk, the location of ice evaporation fronts and dead zones. The observed properties in the infrared range depend on the temperature distribution in the disk. The details of heating and cooling processes determine the development of a number of instabilities, in particular, gravitational, convective, and thermal instabilities, see~e.g.~\cite{2022arXiv220107262A,2022arXiv221013314B}.
In this regard, the problem of describing the self-consistent thermal structure of the disk in numerical simulations can hardly be overestimated.

The main mechanisms of disk heating are the absorption of stellar and interstellar radiation, dissipative processes associated with turbulence and the presence of a magnetic field, and the work of gas pressure forces. The cooling of the disk is mainly caused by diffusion and release of infrared radiation. The combination of these processes leads to the formation of a complex temperature disk structure with vertical stratification and radial gradients. The main source of absorption in the disk is dust, so it is important to know the parameters of the dust itself and how they change during evolution. At relatively small distances from the star, the temperature in the disk can be high enough for the evaporation of dust. Under such conditions, gas becomes the main source of absorption of stellar radiation and, at the same time, the main coolant. The opacity of the gas, in turn, is provided by a wide variety of absorption and emission processes (vibrational, rotational and electronic transitions of various atoms and molecules).

The strong dependence of the gas absorption coefficient on temperature, in particular, due to the ionization of hydrogen, leads to conditions for the development of thermal instability. Under the conditions of circumstellar disks, thermal instability appears when several values of the equilibrium temperature are possible at a fixed surface density, i.e. the disk can be in thermal and hydrostatic equilibrium in one of two phases, which can be called a cold and a hot phase. Assuming that temperature affects the turbulent viscosity of the disk, thermal instability leads to dynamic (viscous) instability of accretion disks. The thermal instability of accretion disks is widely used to explain the activity of dwarf novae, X-ray novae, and other cataclysmic variables, see reviews by~\cite{2020AdSpR..66.1004H,2001NewAR..45..449L}.

Study of thermal structure and thermal instability in
protoplanetary disks is important for understanding the nature of FU~Ori and EX~Lupi type young stellar objects. These objects are outbursting low-mass protostars characterized by a sharp increase in luminosity by tens and hundreds of times \citep{2014prpl.conf..387A,2018ApJ...861..145C}. The nature of these outbursts is still not clear, but the explanation of this phenomenon is necessary for any self-consistent theory of star formation. Despite the fact that the number of known FU~Ori and EX~Lupi type objects is estimated at several dozen \citep{2021A&A...647A..44V}, recent observations of young stellar objects in the optical and near infrared ranges have shown that most young protostars exhibit radiation variability on time scales from several months to several years \citep{2016ApJ...833..104F,2017MNRAS.465.3011C,2017MNRAS.465.3889R}. While short luminosity outbursts in EX~Lupi type objects can be explained by the interaction of the inner boundary of the protoplanetary disk with the protostellar magnetosphere \citep{2012MNRAS.420..416D}, longer FU~Ori type outbursts are most likely caused by processes occurring directly in the protoplanetary disk \citep{2007AstL...33..755K}. The latter may include both magneto-rotational instability \citep{2001MNRAS.324..705A}, gravitational  fragmentation of the disk \citep{2015ApJ...805..115V}, and thermal instability in the inner disk \citep{1994ApJ...427..987B}. These instabilities are essentially sensitive to the thermal structure of the protoplanetary disk, which makes the study of thermal processes an important astrophysical problem.

Thermal instability in a protoplanetary disk in the context to explain the phenomenon of FU~Ori type objects has been modeled in a number of studies. In particular, the work by~\cite{1993PASJ...45..715K} presented periodic luminosity curves obtained in a one-dimensional disk model. The key elements of their model are the simple functions of disk heating and cooling, the postulated switching of the viscosity coefficient between ionized and neutral gas, and convective heat transfer in the vertical direction. \cite{1994ApJ...427..987B} studied the influence of the model parameters on the characteristics of the emerging periodic accretion regime in more detail, adopting similar one-dimensional disk model. \cite{1999ApJ...518..833K} presented the results of two-dimensional hydrodynamic simulations of thermal instability and investigated the transition from the quiet (cold) to active (hot) phase of the evolution of the inner part of the disk. Two-dimensional hydrodynamic calculations of the disk evolution with magneto-rotational and gravitational instabilities by~\cite{1993PASJ...45..715K} also indicated the development of thermal instability in the inner regions of the disk, but the their authors concluded that thermal instability alone is insufficient to provide the  outbursts.

The purpose of this work is to calculate and analyze the distribution of the midplane temperature of the protoplanetary disk based on a model with a more detailed description of a number of physical processes. In particular, we use realistic absorption coefficients by gas and dust, and explicitly include
dust evaporation. In addition, along with viscous heating of the disk, we take into account heating by stellar radiation, while the heating and cooling functions are appropriate for arbitrary optical thicknesses. Based on the presented semi-analytical model, we study the possibility of developing thermal instability in protoplanetary disks.

\section{Model description}

We consider a stationary axially symmetric Keplerian circumstellar disk with the following radial distribution of gas surface density
\begin{equation}
\Sigma^\text{gas}=\Sigma^\text{gas}_\text{au} \left(\dfrac{R}{R_\text{au}}\right)^{-1},
\end{equation}
where surface density is measured from the midplane, i.e., $\Sigma^\text{gas} = \int\limits_{0}^{\infty}\rho(R,z)dz$, $\Sigma^\text{gas}_\text{au}$ is surface density at $R_\text{au}$=1~au, $R$ is the distance between the disk element and the star. The inner and outer disk edges are chosen to be 0.1 and 100~au, respectively. The mass and luminosity of the central star are equal to those of the Sun. In the general case, two sources of disk heating are taken into account: stellar radiation and viscous heating. Stellar radiation is assumed to be blackbody with a temperature of 6000~K. The viscous heating of the disk is calculated in the stationary accretion approximation with a given accretion rate $\dot{M}$. The disk is cooled via release of thermal radiation.

\subsection{Method for calculating midplane temperature}

The midplane temperature $T_\text{mid}$ at each distance $R$ from the star is found from the balance between heating and cooling:
\begin{equation}
\Lambda_\text{IR}=\Gamma_\text{star}+\Gamma_\text{vis},
\label{Tmid}
\end{equation}
where $\Lambda_\text{IR}$ is the cooling rate of the disk midplane layers due to escape of IR radiation, $\Gamma_\text{star}$ is the heating rate of the midplane layers by stellar radiation, $\Gamma_\text{vis}$ is the heating rate due to viscous dissipation:
\begin{flalign}
 &\Lambda_\text{IR}=
   \frac{4\tau_\text{P} \sigma T_\text{mid}^4}{1+2\tau_\text{P}\left(1+\dfrac{3}{4}\tau_\text{R}\right)} 
   \label{Tmid1}\\
 &\Gamma_\text{star} = \nonumber\\
 &=\frac{\mu F_0\tau_\text{P}\left[
   2\left(1-e^{-\tau_\text{uv}}\right) + 3\mu \dfrac{\tau_\text{R}}{\tau_\text{uv}}
   \left(1-e^{-\tau_\text{uv}}-\tau_\text{uv}e^{-\tau_\text{uv}}\right)
   +\frac{\tau_\text{uv}}{\tau_\text{P}} e^{-\tau_\text{uv}}
   \right]}{1+2\tau_\text{P}\left(1+\dfrac{3}{4}\tau_\text{R}\right)}
   \label{Tmid2}\\
 &\Gamma_\text{vis}=\frac{3}{8\pi}\frac{GM\dot{M}}{R^3}.
   \label{Tmid3}
\end{flalign}
\\
\noindent The detailed derivation of the above functions is presented in the Appendix. In these equations, $F_0$ is the radiation flux from the star reaching the disk surface, $\mu$ is the cosine of the angle between the direction to the star and the normal to the disk surface (chosen to be 0.05), $\tau_\text{uv} $ is optical depth to stellar radiation, $\tau_\text{P}$ and $\tau_\text{R}$ are Planck and Rosseland mean optical depths to disk thermal radiation, $\sigma$ is the Stefan-Boltzmann constant, $G$ is the gravitational constant, $M$ is the mass of the central star, $\dot{M}$ is the accretion rate over the disk. Optical depths are defined as follows:
\begin{align}
&\tau_\text{uv}=\dfrac{1}{\mu}(\kappa_\text{F}^\text{gas}\Sigma^\text{gas}
+\kappa_\text{F}^\text{dust}\Sigma^\text{dust}) \label{tau_uv}\\
&\tau_\text{P}=\kappa_\text{P}^\text{gas}\Sigma^\text{gas}
+\kappa_\text{P}^\text{dust}\Sigma^\text{dust} \label{tau_P}\\
&\tau_\text{R}=\kappa_\text{R}^\text{gas}\Sigma^\text{gas}
+\kappa_\text{R}^\text{dust}\Sigma^\text{dust}, \label{eqRos}
\end{align}
where $\Sigma^\text{gas}$ and $\Sigma^\text{dust}$ are gas and dust surface densities, $\kappa_\text{F}^\text{gas}$, $\kappa_\text{P}^\text{gas}$, and $\kappa_\text{R}^\text{gas}$ are
two-temperature mean, Planck mean, and Rosseland mean opacity coefficients for gas, respectively, $\kappa_\text{F}^\text{dust}$, $\kappa_\text{P}^\text{dust}$, and $\kappa_\text{R}^\text{dust}$ are dust
opacity coefficients. Opacity coefficients are generally functions of density and temperature. Note that the expression~\eqref{eqRos} is valid in the gray body approximation, but is incorrect in the general case, i.e. Rosseland mean optical depths are not additive. This is due to the non-linear nature of the averaging procedure. However, we use this formula because it provides acceptable accuracy and significantly simplifies the subsequent analysis.

Equations~\eqref{Tmid1}--\eqref{Tmid2} are derived under the assumption that the opacity coefficients are constant along the vertical direction. This is obviously a rather rough approximation, since both temperature and density change significantly with height (distance from the midplane). Nevertheless, we use these expressions because they allow to describe the thermal structure of the disk in the first-order approximation. Temperature and density in the midplane are used as arguments for calculating the opacity coefficients. The gas opacity coefficients, as will be shown below, depend on gas density.
At a fixed midplane temperature, the midplane density is found from the condition of hydrostatic equilibrium of a vertically isothermal disk:
\begin{equation}
\rho_\text{mid}= \frac{1}{\sqrt{2\pi}} \frac{2\Sigma^\text{gas}}{H}
\label{static}
\end{equation}
where $H = c_s/\Omega$ is the disc scale height, 
$c_s = \sqrt{k_\text{B}T_\text{mid}/m}$ is isothermal sound speed,
$m$ is molecular mass, $k_\text{B}$ is Boltzmann constant, 
$\Omega =\sqrt{GM/R^3}$ is Keplerian angular velocity at radius $R$.

Since the opacity coefficients are functions of temperature, the equation~\eqref{Tmid} is a non-linear equation with respect to $T_\text{mid}$. We find its solution graphically by calculating the sign of the difference between the left and right sides of equation~\eqref{Tmid} for a sequence of trial values of $T_\text{mid}$. The areas of sign change of this difference correspond to the roots of the equation.

\subsection{Gas absorption coefficients}

\begin{figure*}
\includegraphics[width=0.8\columnwidth]{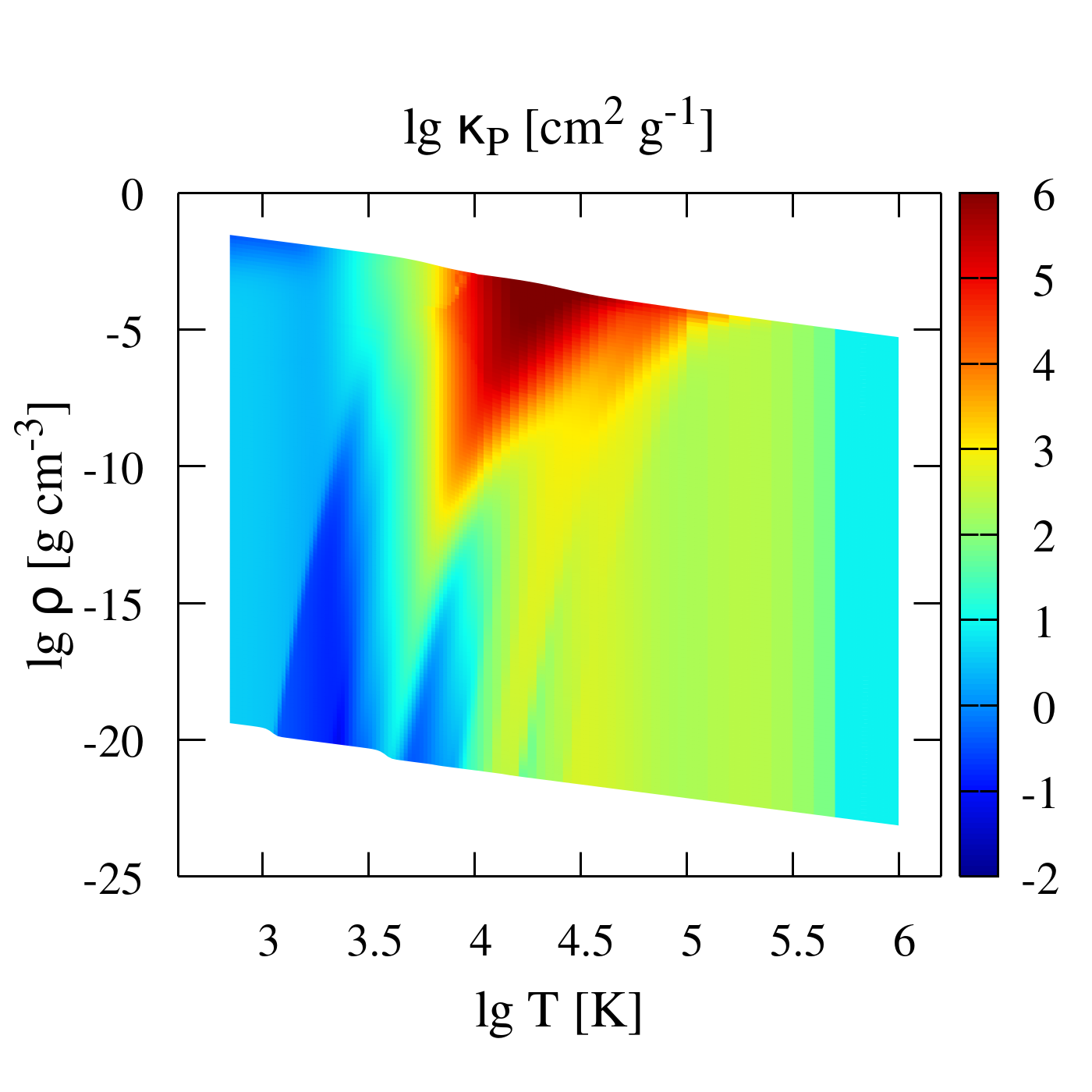}
\includegraphics[width=0.8\columnwidth]{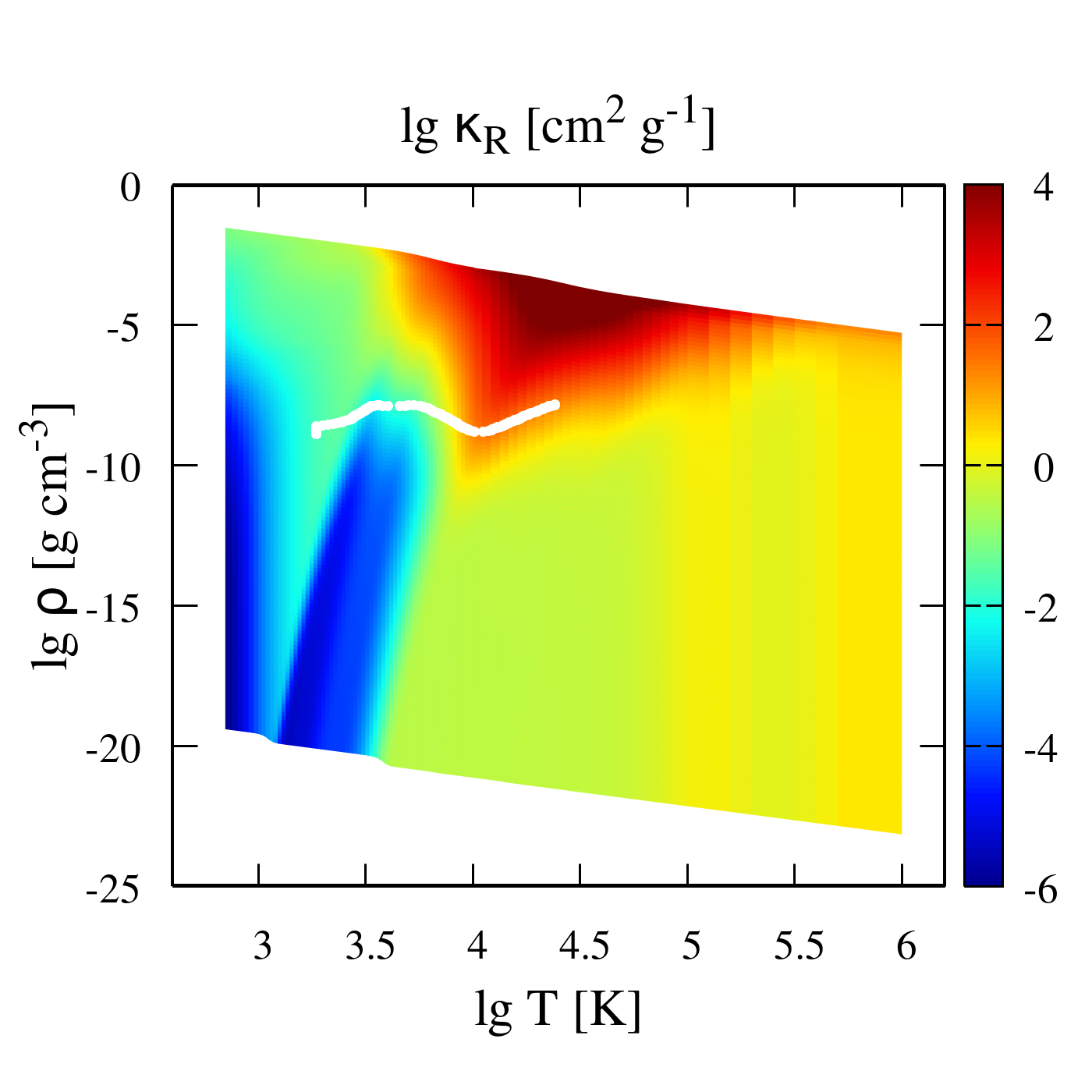}
\includegraphics[width=0.8\columnwidth]{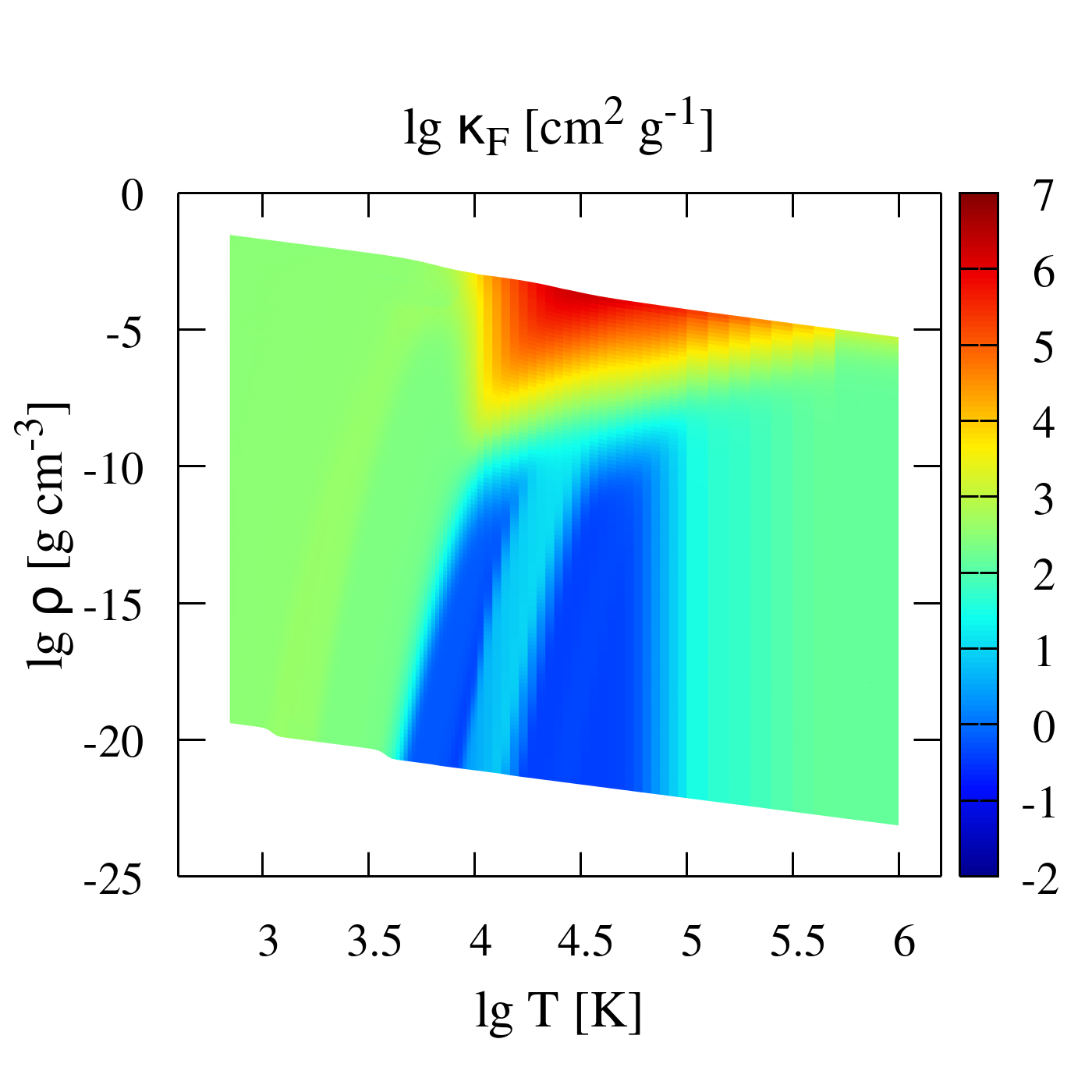}
\caption{Planck mean (top left panel), Rosseland mean (top right panel), and two-temperature mean (bottom panel) gas opacity coefficients from~\citet{2014A&A...568A..91M}. The white curve in the top right panel corresponds to the equilibrium parameters of the disk in the instability region from the M5 model, see Fig.~\ref{results}.}
\label{gas-opacity}
\end{figure*}

Gas opacity coefficients 
$\kappa_\text{P}^\text{gas}$, 
$\kappa_\text{R}^\text{gas}$, 
$\kappa_\text{F}^\text{gas}$ 
as functions of temperature and gas density (see Fig.~\ref{gas-opacity}) are adopted from~\cite{2014A&A...568A..91M}\footnote{\url{https://vizier.cds.unistra.fr/viz-bin/VizieR?-source=J/A+A/568/A91}}. To calculate these coefficients, \cite{2014A&A...568A..91M} used the DFSYNTHE code~\citep{2005MSAIS...8...34C,1970SAOSR.309.....K}, originally oriented for calculation of stellar atmospheres. The coefficients we use correspond to the solar metallicity, $\kappa_\text{F}$ is taken for a stellar temperature of 6000~K. We note that at relatively low temperatures ($T<1000$~K) the main contribution to $\kappa_\text{P}^\text{gas}$ and $\kappa_\text{R}^\text{gas}$ comes from the absorption lines of various molecules. At higher temperatures ($T>3000$~K), absorption by atomic hydrogen, the H$^-$ ion, various metals, and other processes become important. Note also that the distribution of $\kappa_\text{F}^\text{gas}(T,\rho)$ differs significantly from the distributions of $\kappa_\text{P}^\text{gas}(T,\rho) $ and $\kappa_\text{R}^\text{gas}(T,\rho)$.

When calculating the thermal structure of the disk for arbitrary $T_\text{mid}$ and $\rho_\text{mid}$, we use linear interpolation of the coefficients between neighboring nodes of the input grid and extrapolation of the opacity coefficients by edge values in case of going beyond the original grid in terms of temperature and density.

\subsection{Dust absorption coefficients and evaporation model}

\begin{figure*}
\includegraphics[width=1\columnwidth]{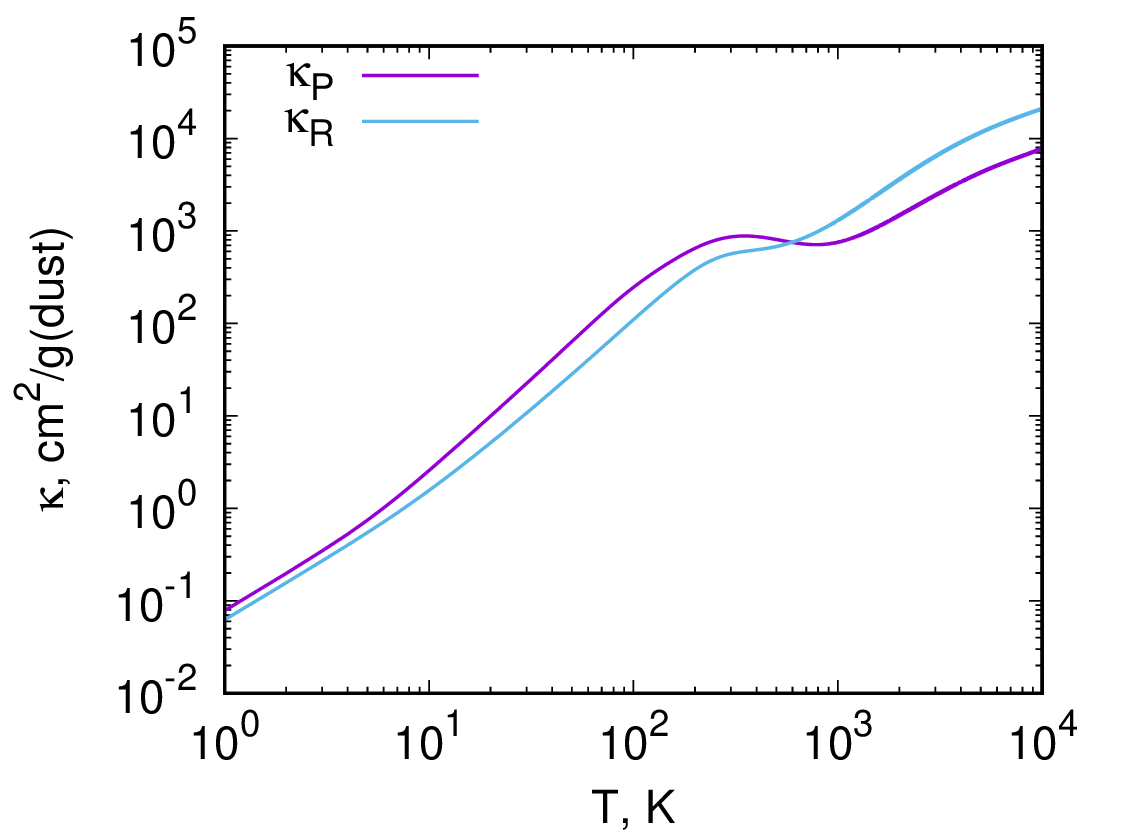}
\includegraphics[width=1\columnwidth]{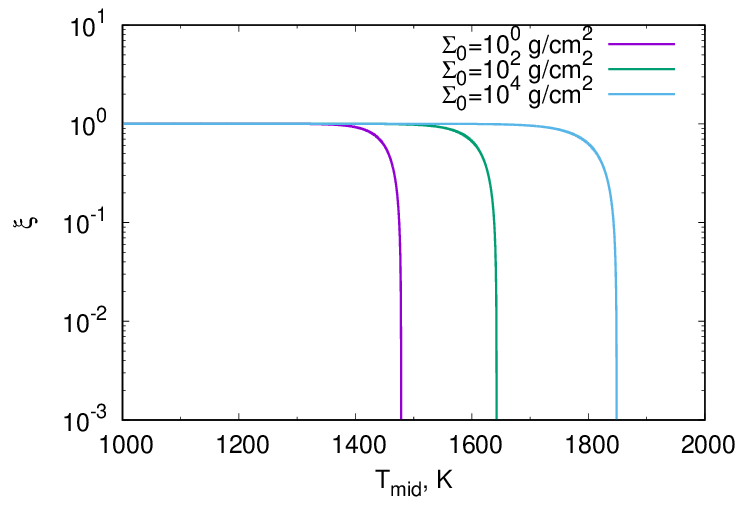}
\caption{Left panel: Planck and Rosseland mean dust opacity versus temperature. Right panel: fraction of unevaporated dust as a function of midplane temperature for three values of the disk surface density at 1~au .}
\label{dust-opacity}
\end{figure*}

The adopted Planck and Rosseland mean opacitis as functions of temrerature are shown in the left panel of Fig.~\ref{dust-opacity}. We calculate these coefficients  from the frequency-dependent absorption and scattering coefficients for spherical silicate dust grains. We chose forsterite Mg$_2$SiO$_4$ as a model of silicate dust. The spectral absorption and scattering coefficients themselves were calculated using the Mie theory, with the size distribution of dust grains taken as a power-law $n(a)\propto a^{-3.5}$ with the minimum and maximum dust grain radii $a_{\rm min}=5 \times 10^{-7}$~cm and $a_{\rm max}=10^{-4}$~cm. {Note that the opacity is shown in Fig.~\ref{dust-opacity} up to $10^4$~K, although the dust should evaporate at much lower temperatures (see below). However, the calculation of the dust opacity coefficients (per unit mass of dust) and the fraction of evaporated dust are independent procedures, the combination of which will give the optical depth necessary for modeling the thermal structure based on the relations \eqref{tau_uv}--\eqref{eqRos}.}

At high temperatures the dust evaporates. We account for this process by calculating the surface density of the dust as follows:
\begin{equation}
\Sigma^\text{dust} = \xi(T_\text{mid})\, \mu_\text{dg}\, \Sigma^\text{gas},
\end{equation}
where $\mu_\text{dg}=0.01$ is the dust-to-gas mass ratio we assume in the absence of evaporation, $\xi(T)$ is the fraction of unevaporated dust at temperature $T$. Thermal evaporation of forsterite is a complex process, it is accompanied by the formation of a number of components in the gas phase, such as Si, O, Mg, MgO, O$_2$, SiO, SiO$_2$. We calculate $\xi(T)$ assuming that dust is in thermodynamic equilibrium between the solid and the gas phases. In this approximation, the coexistence of phases is possible provided that the partial pressures of gases from the dust components are equal to the pressures of their saturated vapors at a given temperature. The calculation of equilibrium partial pressures is carried out by the methods of chemical thermodynamics and is presented, for example, in the paper by~\cite{1996A&A...312..624D}.

We  consider an acceptable simplification that all silicon in the gaseous phase is in the form of a SiO molecule. To calculate the function $\xi(T)$ we use the following expression:
\begin{equation}
\xi(T) = 1-\dfrac{n_\text{vap}}{n_\text{tot}},
\label{eq_xi}
\end{equation}
where $n_\text{vap}$ is number density of silicon (in the form of SiO) in saturated vapor at temperature $T$,
$n_\text{tot}$ is total number density of silicon (in solid and gaseous phases).
Let us express the number density $n_\text{vap}$ in terms of saturated vapor pressure $P_\text{vap}$:
\begin{equation}
n_\text{vap}=\frac{P_\text{vap}}{k_\text{b}T}.
\end{equation}
We take the approximation of $P_\text{vap}$ temperature dependence for silicon oxide vapor from~\cite{1996A&A...312..624D} as:
\begin{equation}
P_\text{vap}=f_\text{a}\, \exp \left(x_1/T + x_2 +x_3 T +x_4 T^2+x_5 T^3\right),
\end{equation}
with the parameters $x_1=-6.28\times 10^4$~K, $x_2=1.80\times 10^1$,
$x_3=3.59\times 10^{-4}$~K$^{-1}$, $x_4=-3.72 \times 10^{-7}$~K$^{-2}$,
$x_5=6.53\times 10^{-11}$~K$^{-3}$, and the coefficient $f_\text{a}=10^6$~dyn cm$^{-2}$ which takes into account the conversion of pressure to CGS units.

The total number density of silicon $n_\text{tot}$ is considered proportional to the number density of the matter in the midplane:

\begin{equation}
n_\text{tot}=\frac{X_\text{Si}\, \rho_\text{mid}}{\mu_\text{Si}\,m_\text{a}},
\end{equation}
where $X_\text{Si}=3.5\times 10^{-5}$ is mass fraction of silicon in the interstellar medium, $\mu_\text{Si}=28$ is atomic weight of silicon,
$m_\text{a}$ is atomic unit of mass. Since $n_\text{tot}$ is determined by the midplane density $\rho_\text{mid}$, which depends on a number of parameters (see  equation~\eqref{static}), then the dependence $\xi(T)$ will be in general unique for each $R$ in the disk.

At a sufficiently high temperature, $n_\text{vap}$ may turn out to be greater than 
$n_\text{tot}$, and $\xi$ becomes formally negative according to the formula~\eqref{eq_xi}. This means that at a given temperature, the silicon available in a given volume is insufficient to saturate the vapors, and phase equilibrium is impossible. In this case, all silicon should pass into the gaseous state, i.e. the dust is completely evaporated. To avoid physical uncertainties in the numerical model, we additionally restrict the smallest value of $\xi$ by the parameter
$\xi_\text{min}=10^{-5}$, i.e. we assume that under the condition $n_\text{vap}>n_\text{tot}$ a small fraction of dust grains does not evaporate.

The right panel of Fig.~\ref{dust-opacity} shows the function $\xi(T_\text{mid})$ for three values of the surface density of the disk at 1~au ($\Sigma=$1, $10^2$, $10^4$~g cm$^{-2}$) around a solar mass star. It can be seen that at low temperatures, the function $\xi(T)$ is close to unity, and when the region of intense evaporation is reached, the function $\xi(T)$ decreases very rapidly.

{The adopted dust model is quite primitive and is used only to study the physical effect of evaporation. In fact, protoplanetary dust can include a carbon component, refractory organic matter, polycyclic aromatic hydrocarbons, ice mantles. The dust particles themselves can have a complex and fractal shape.}

\subsection{Considered models}
The ``M1'' model considers a sparse low-mass ($M_\text{disk}$ $\approx$ $10^{-5} M_\odot$) disk heated only by stellar irradiation, and only takes into account dust opacity, while the the dust is assumed to be non-evaporable. In the ``M2'' model, the surface density of the disk is increased by four orders of magnitude relative to the ``M1'' model, such a massive disk ($M_\text{disk}$ $\approx$ 0.1 $M_\odot$) corresponds to the initial phases of the evolution of a protostellar system. In the ``M3'' model, viscous heating corresponding to the accretion rate of $10^{-6}M_\odot$/yr is added to ``M2'' as an additional source of heating. In the ``M4'' model, the restriction on the impossibility of dust evaporation is removed. Finally, the ``M5'' model takes into account, in addition to everything else, the contribution of the gas to absorption and emission. Designations and parameters of considered models are given in Table~\ref{tab_models}.

\begin{table}
\centering
 \begin{tabular}{c|c|c|c|c|c}
 \hline
Model & Dust & Dust & Gas & $\Sigma_0$, g/cm$^2$ & $\dot{M}$, $M_\odot$/yr \\
      & Opacity & Evaporation & Opacity & & \\
\hline
M1 & $+$ & $-$ & $-$ & $0.1$ & 0 \\
M2 & $+$ & $-$ & $-$ & $1000$ & 0 \\
M3 & $+$ & $-$ & $-$ & $1000$ & $10^{-6}$ \\
M4 & $+$ & $+$ & $-$ & $1000$ & $10^{-6}$ \\
M5 & $+$ & $+$ & $+$ & $1000$ & $10^{-6}$ \\
 \hline
 \end{tabular}
\caption{Considered disk models and their parameters}
\label{tab_models}
\end{table}

\section{Simulation results}

\begin{figure*}
\includegraphics[width=0.90\columnwidth]{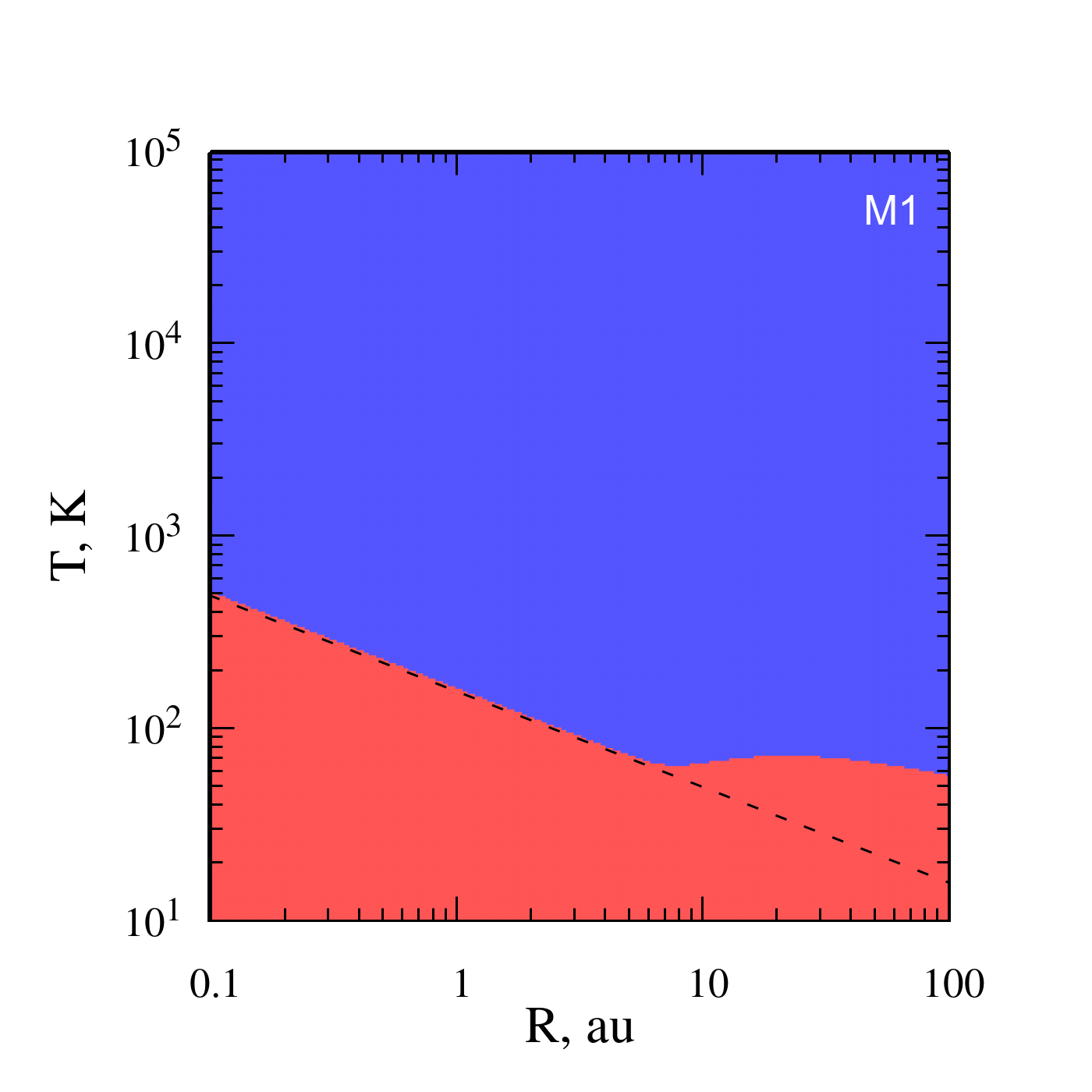}
\includegraphics[width=0.90\columnwidth]{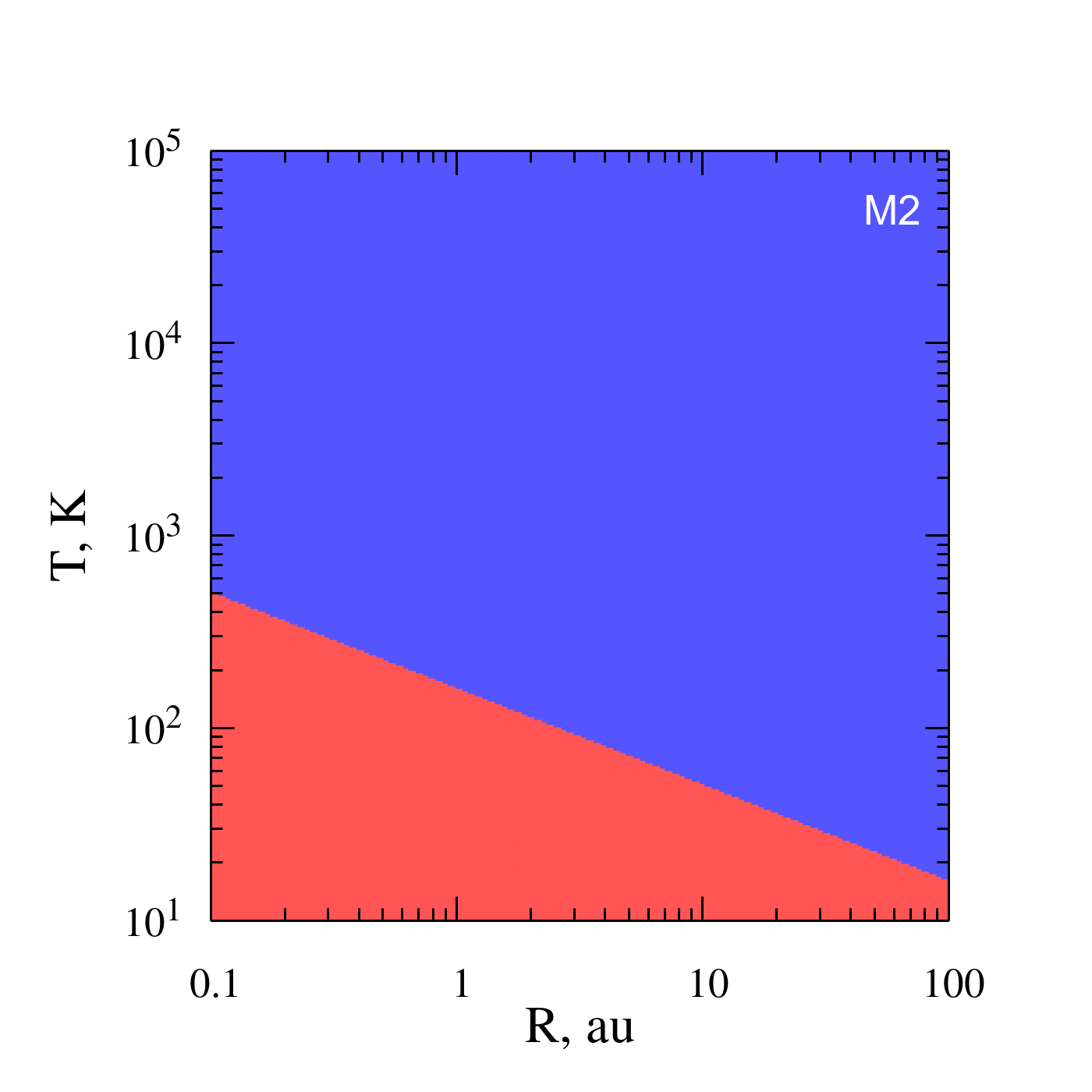}\\
\includegraphics[width=0.90\columnwidth]{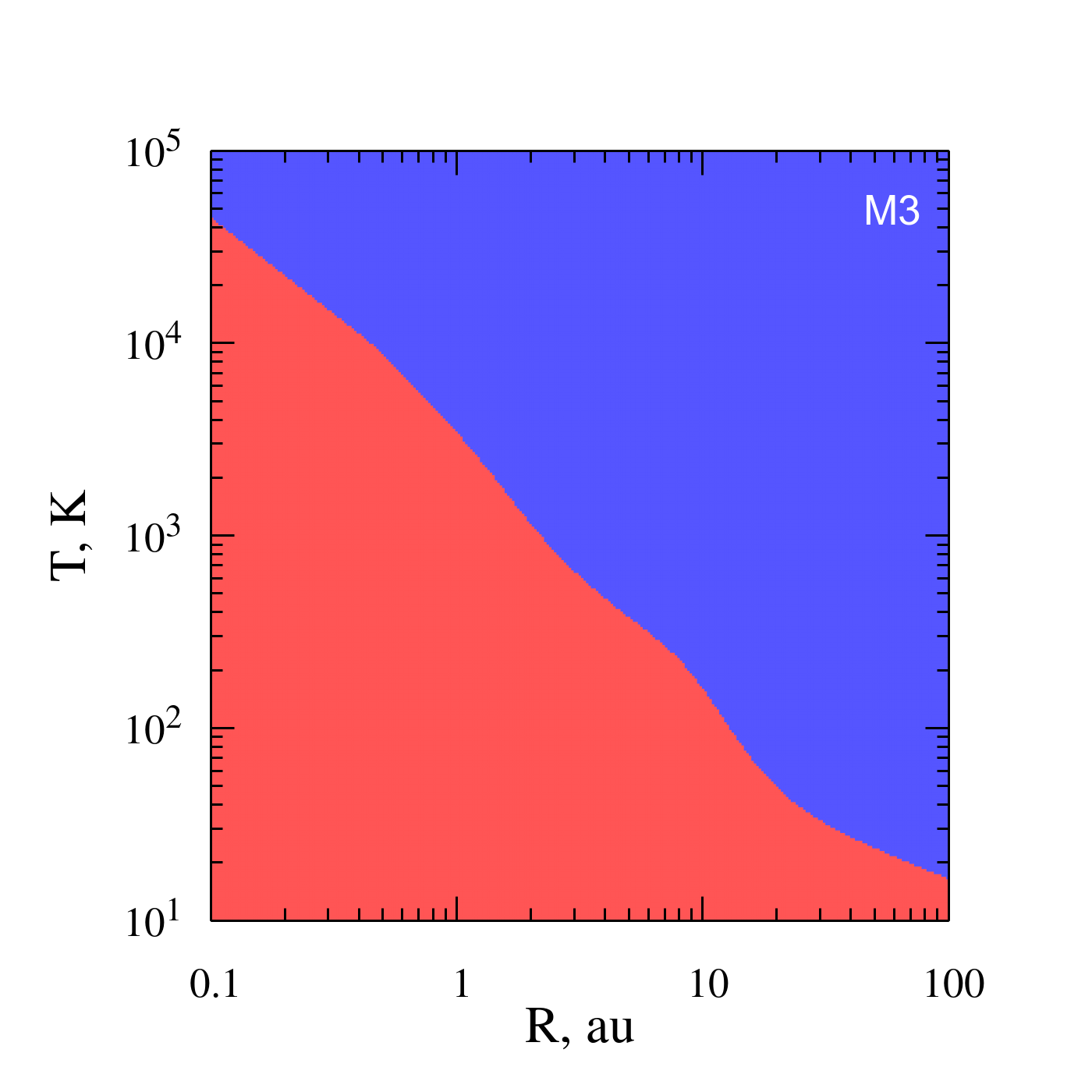}
\includegraphics[width=0.90\columnwidth]{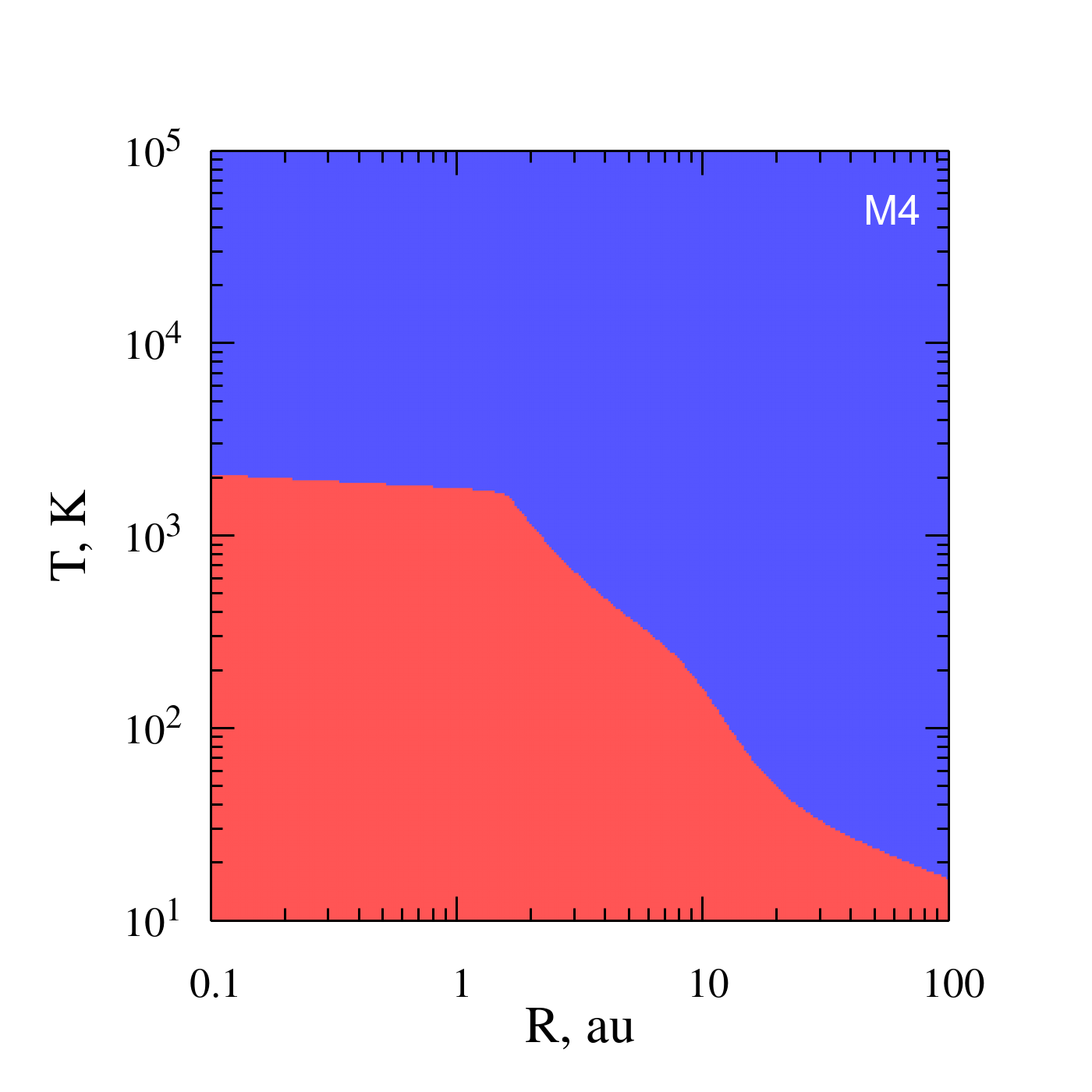}\\
\includegraphics[width=0.90\columnwidth]{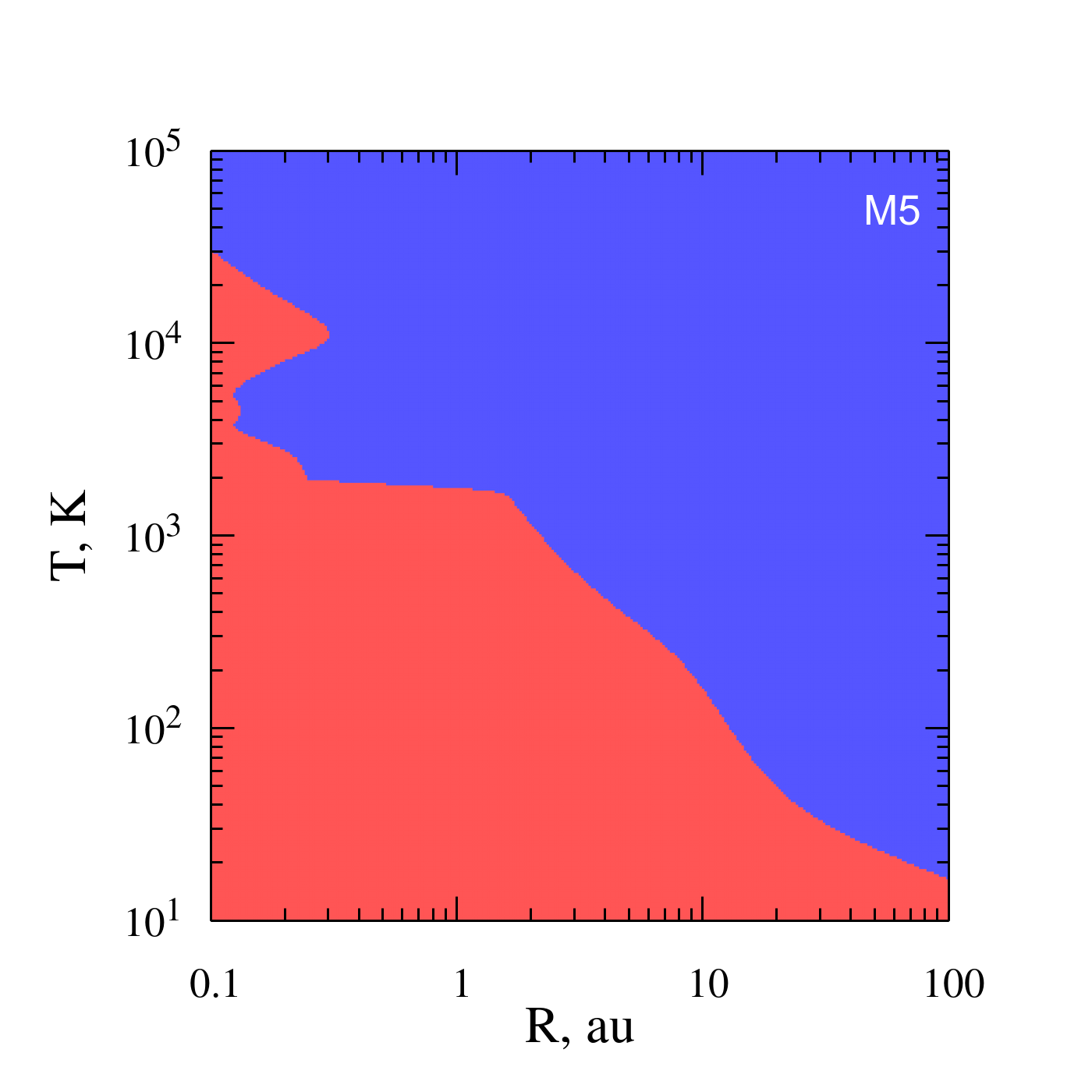}
\includegraphics[width=0.90\columnwidth]{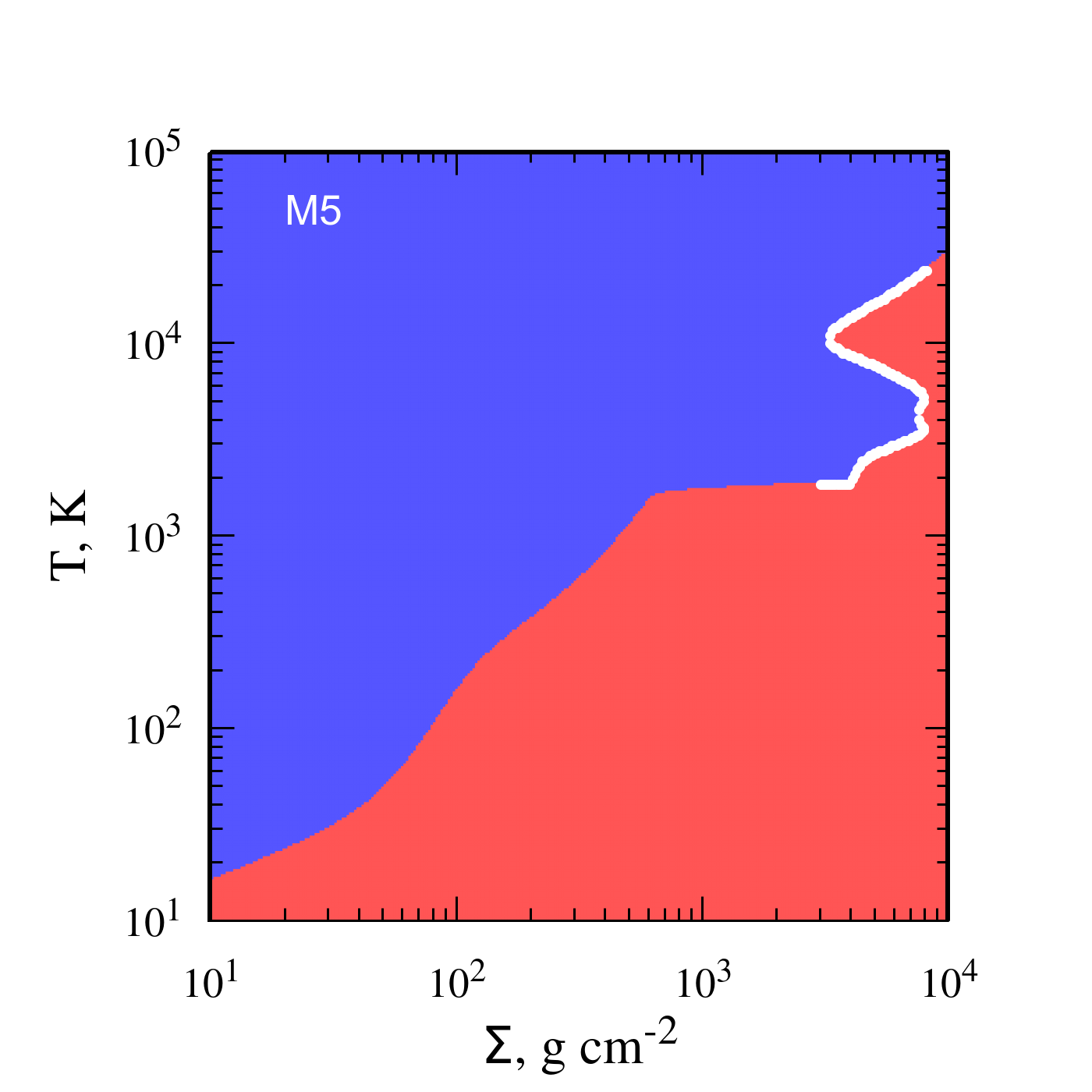}
\caption{Calculated disk thermal structure for models M1--M5. The equilibrium temperature lays at the border between the red and blue regions, the color is determined by the sign of the difference between the right and left parts of the equation~\eqref{Tmid}. The dashed line shows the distribution according to equation~\eqref{lim2b}. The white curve highlights the multi-valued solution that is responsible for the instability.}
\label{results}
\end{figure*}

Fig.~\ref{results} shows the results of calculating the temperature distributions for the considered models. The equilibrium temperature lays at the border between the red and blue regions, the color is determined by the sign of the difference between the right and left parts of the equation~\eqref{Tmid}.

In the M1 model, the temperature gradually decreases with distance {up to a radius of 7~AU,} after which a ``hump'' forms on the distribution. The monotonic inner portion of the distribution corresponds to the region of the disk opaque to stellar radiation and is well described by the relation~\eqref{lim2}:
\begin{equation}
aT^4_\text{mid}  = \dfrac{2\mu F_0}{c},
\label{lim2b}
\end{equation}
which depends only on the parameters of the stellar radiation flux. Dependence~\eqref{lim2b} is shown in the figure by a dashed line. {The disk in the M1 model is so rarefied that the outer parts of this disk ($R>7$~AU) are transparent to stellar radiation. The formation of a temperature hump is associated with a smooth transition to the limit~\eqref{lim1} for a disk that is optically thin to stellar radiation:
\begin{equation}
aT^4_\text{mid} = \frac{\kappa_\text{F}}{\kappa_\text{P}}\frac{F_0}{c},
\label{lim1b}
\end{equation}
where the thermal structure depends not only on the radiation flux, but also on the opacity ratio $\kappa_\text{F}/\kappa_\text{P}$.}

The optical depth to stellar radiation in the M2 massive disk model is large for the entire disk, so the distribution of the midplane temperature is monotonous and is determined only by the dilution of stellar radiation in accordance with the formula~\eqref{lim2b}.

In the M3 model of a massive disk with viscous heating, the temperature distribution depends not only on the stellar flux parameters, but also on the dissipation rate and the opacity of the medium. The thermal structure of such a disk is described by the expression~\eqref{lim3}:
\begin{equation}
aT^4_\text{mid} = \dfrac{2\mu F_0}{c} +\frac{\Gamma_\text{vis}}{c}\left[\frac{
1+2\tau_\text{P}(1+\frac{3}{4}\tau_\text{R})}{\tau_\text{P}}
\right].
\label{lim3b}
\end{equation}
With the adopted model parameters, the temperature changes from $4\times 10^4$~K to 20~K when moving from the inner to the outer boundary of the disk. The uneven distribution of the equilibrium temperature is related to the features of the $\kappa_\text{P}^\text{dust}(T)$ and $\kappa_\text{R}^\text{dust}(T)$ dependences. In particular, these features are associated with the nonmonotonic behavior of the absorption coefficient of silicate dust in the vicinity of 10 microns. Note that the temperature inside 2~au turns out to be much higher than the dust evaporation temperature (which does not exceed 2000~K, see Fig.~\ref{dust-opacity}). Therefore, the M3 model is certainly physically inconsistent, but we present it for methodological purposes to show what the considered approximations lead to when calculating the thermal structure of a disk.

In the M4 model, which takes into account both viscous heating and dust evaporation, the temperature in the inner region of the disk ($R<2$~au) is significantly lower than in the ``M3'' model. The temperature in this region is $T_\text{mid}\approx$2000~K. When such temperatures are reached in the specified region, the abundance of unevaporated dust becomes equal to the minimum value postulated in the model $\xi_\text{min}=10^{-5}$. The value of $10^{-5}$ was chosen in a way that the corresponding optical depths $\tau_\text{P}$ and $\tau_\text{R}$ in this region are close to unity, thereby ensuring the maximum cooling rate ~$\Lambda_\text{IR}$. In this case, the cooling rate~$\Lambda_\text{IR}$ fully compensates for the heating rate~$\Gamma_\text{star}$ and $\Gamma_\text{vis}$, providing the equilibrium value of $T_\text{mid}$. A plateau with a slightly lower temperature $T_\text{mid}\approx 1600$~K is also described in the paper by~\cite{1999ApJ...527..893D} (see their Fig.~1 and description on page .~895, third paragraph of section 2.2). \cite{1999ApJ...527..893D} also take dust sublimation into account, while their gas opacity is small for these conditions, which makes their model conceptually close to our M4 model. Differences in plateau temperatures are probably due to different models of dust evaporation.

In the M5 model, which takes into account the contribution of the gas to the opacity of the medium, the optical depths in the inner part of the disk increase significantly, which generally leads to an increase in temperature compared to the M4 model. The key feature of this model is that the equilibrium temperature {within 0.15--0.3~au} has multiple solutions, which is caused  by the strong dependence of gas opacity on temperature. The red color on the distributions corresponds to heating dominating over cooling, while the blue color corresponds to the predominance of cooling. With this in mind, the arrival at equilibrium can be considered in this diagram as an upward movement (heating leads to an increase in temperature) in the red region and a downward movement (cooling leads to a decrease in temperature) in the blue region to the boundary of the regions. Thus {within 0.15--0.3~AU} stable equilibrium is possible at temperatures of 2--3 thousand~K or 10--20 thousand~K. Taking into account the ambiguity of the solution for the equilibrium temperature, it can be expected that in the inner zone of the considered disk, the conditions for thermal instability are fulfilled. The actual temperature will be determined by the side of the distribution from which the disk comes to the equilibrium.

It is also useful to represent the results obtained for the M5 model as the function of surface density $T_\text{mid}(\Sigma)$ shown in the lower right panel of Fig.~\ref{results}. The resulting S-shaped distribution of the equilibrium temperature $T(\Sigma)$ inevitably leads to an analogy with the shape of the $\Sigma-T_\text{eff}$ dependence for classical thermal instability in accretion disks of cataclysmic variables~\citep{2001NewAR..45..449L}, where the formation of an equilibrium temperature bend is associated with hydrogen ionization, which leads to a strong dependence of opacity on temperature. In Fig.~\ref{gas-opacity} the white curve shows the values of midplane density and temperature corresponding to the region of thermal instability. One can see that the white curve crosses the area of strong gradients in the opacity distribution. This result allows us to say that when the opacity of gas in the inner regions of protoplanetary disks is taken into account, the development of instability is possible. This, among other things, can lead to the formation of morphological features in the inner regions of the disks and/or a nonstationary (episodic) character of accretion. The episodic accretion will naturally appear in the model of viscous disk evolution if the $\alpha$-parameterization of turbulent viscosity is used. With such a parameterization, there is a positive feedback between the disk temperature and the accretion rate, which, in the presence of a jump in the equilibrium temperature, leads to an accumulative mode of disk evolution, followed by a rapid accretion of disk matter onto the star. An example of such a model is discussed in the next section.

\section{Evolutionary disk model}

\begin{figure*}
\includegraphics[width=0.66\columnwidth]{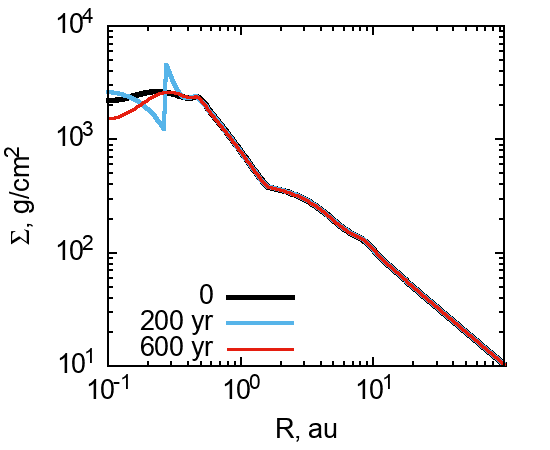}
\includegraphics[width=0.66\columnwidth]{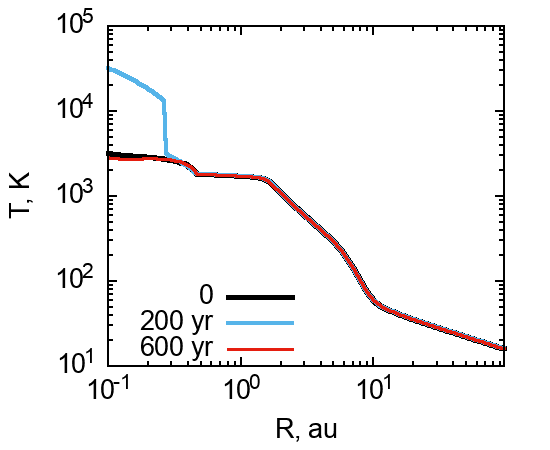}
\includegraphics[width=0.66\columnwidth]{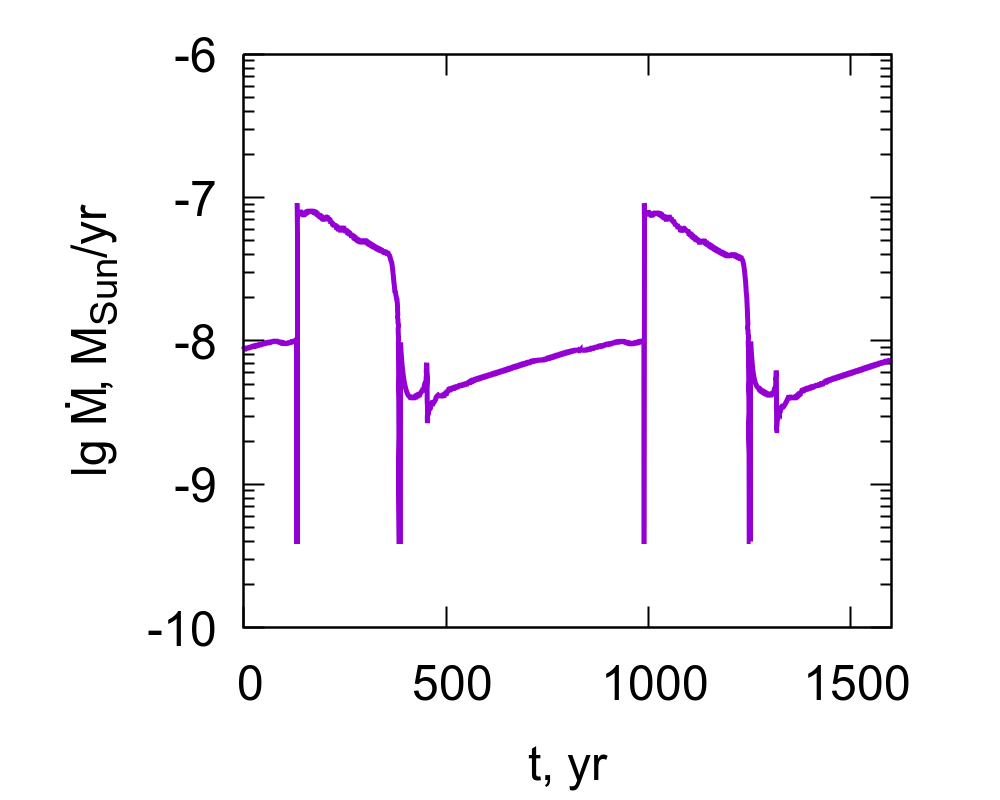}
\caption{Results of the calculation of the viscous disk evolution. Left panel: surface density distributions for {three moments}. Middle panel: midplane temperature distributions for the same time moments. Right panel: disk-to-stellar accretion rate change over time. {The zero moment of time is chosen arbitrarily and corresponds to 15.4 thousand years of evolution from the initial state}.}
\label{evolution}
\end{figure*}

To illustrate possible influence of the discussed instability on the nature of accretion in a protoplanetary disk, we consider a model of a viscous disk, the evolution of which is described using the following system of equations:
\begin{eqnarray}
&&\frac{\partial \Sigma}{\partial t} = 
\frac{3}{R}\frac{\partial}{\partial R} \left(R^{1/2} \frac{\partial }{\partial R}\left(R^{1/2} \nu \Sigma \right)\right)
\label{pringle1} \\
&&\frac{\partial \varepsilon}{\partial t} = \Gamma_\text{star} + \Gamma_\text{vis} - \Lambda_\text{IR} + 
\frac{3}{R}\frac{\partial}{\partial R} \left(R^{1/2} \frac{\partial }{\partial R}\left(R^{1/2} \nu \varepsilon \right)\right),
\label{pringle2}
\end{eqnarray}
where $\varepsilon = c_\text{v}T_\text{mid}\Sigma$ is thermal energy per unit disk surface, $c_\text{v} = \dfrac{k_\text{B}}{m(\gamma-1)}$, $\gamma=7/5$ is adiabatic exponent, $\Gamma_\text{star}$, $\Gamma_\text{vis}$ are the rates of disk heating by stellar radiation and viscous dissipation, calculated by the equations~\eqref{Tmid2} and \eqref{Tmid3}, respectively, $\Lambda_\text{IR}$ is cooling rate due to IR emission \eqref{Tmid1}, $\nu$ is turbulent viscosity coefficient. The equation~\eqref{pringle1} for surface density is the classical Pringle equation.

Equation~\eqref{pringle2} for the evolution of thermal energy along with the rates of heating and cooling includes radial transfer of thermal energy (the last term on the right side). The transfer of thermal energy is treated here similarly to the transfer of individual components in the diffusion accretion-decretion disk approximation. In this approximation, it is assumed that turbulence leads to efficient mixing of matter, i.e. to turbulent diffusion. As a result, the dynamics of all individual components of matter (for example, impurities) is described by the same diffusion equation, similar to the Pringle equation, see equation (13) in~\cite{2007A&A...471..833P}. Assuming that thermal energy is inextricably linked with the matter itself, the equation for its transfer in this approximation is similar to the Pringle equation. At zero rates of heating and cooling, such a system of equations leads to the fact that the initially isothermal disk will remain so in the course of evolution. We note that for the case of a purely viscous disk, the spatial dynamics of individual components (and thermal energy) must be described using the advection equation. The relationship between the viscosity and diffusion coefficients (Schmidt number) in protoplanetary disks is a debatable issue, see e.g.~\cite{2007A&A...471..833P}, and therefore the choice in favor of one or another approximation is not obvious. Note, however, that the use of the diffusion approximation to calculate the thermal energy transfer greatly simplifies the numerical model and, at the same time, reflects the main features of the spatial redistribution of energy.

The viscosity coefficient is given in the framework of the classical $\alpha$ parameterization of~\cite{1973A&A....24..337S}: $\nu=\alpha c_\text{s} H$,
where $c_\text{s}$ is speed of sound in the midplane,
$H$ is typical disk scale height. The scale height is calculated from the vertical hydrostatic equilibrium condition: $H/R=c_\text{s}/v_\text{k}$. We use the constant value of $\alpha=10^{-3}$. The surface density distribution from the M5 model is taken as the initial one.

The integration of equations~\eqref{pringle1}-- \eqref{pringle2} at each time step is carried out in two stages. At the first stage, we solve the problem of diffusion of surface density and thermal energy. At the second stage, the change in thermal energy due to heating and cooling sources is calculated. For both stages, implicit methods are used (the tridiagonal matrix algorithm for solving a system of linear algebraic equations when calculating diffusion and the bisection method for updating the temperature due to the heating and cooling, respectively), which allows one to significantly weaken the constraint on the time step.

Fig.~\ref{evolution} shows the calculation results of the considered disk evolution model {after 15.4~thousand years} from the initial distribution. In the inner region of the disk {($R<0.5$~au)}, the disk structure is periodically rearranged: when a sufficient amount of matter is accumulated, the IR radiation is ``locked'', which leads to an increase in temperature and a transition to the upper branch of the quasi-equilibrium solution. High temperature, in turn, leads to increased viscous heating, which maintains a high-energy regime until the moment when a significant part of the matter from the inner region accretes onto the star as a result of increased viscosity. This regime leads to an episodic accretion pattern with a {period of about 1000 years}, which, however, will change as the disk is depleted. We note that the characteristics of flare activity in this model strongly depends on the viscosity coefficient. The maximum accretion rate decreases with $\alpha$, while the flare period increases, and the flare shape also changes.

\section{Conclusions}

The model of the thermal structure of a protoplanetary disk presented in this article is mainly illustrative due to a number of rather rough approximations. Its main goal was to demonstrate that conditions for thermal instability can indeed be fulfilled in a protoplanetary disk. Undoubtedly, a full study of the evolution of the protoplanetary disk, taking into account dust evaporation and using realistic gas extinction coefficients, should be carried out on the basis of a more consistent dynamical model. In such a model it is particularly necessary to abandon the approximation of thermodynamic equilibrium between the solid and gaseous phases when calculating the fraction of evaporated dust. Instead, it should be taken into account that the dust evaporates and condenses in a finite time, and the characteristic times of these processes can differ significantly. The dynamic model must take into account not only the motion of the gas, but also the drift, settling, and growth of dust grains, since these processes have a strong influence on the thermal and spatial structure of the disk. Nevertheless, the expressions for the rates of heating and cooling presented in this paper, together with the formalism of calculating optical depths taking into account gas opacity and dust evaporation, can be used for more detailed modeling of the long-term evolution of the disk, for example, within the model of a viscous self-gravitating disk from the studies by~\cite{2015ApJ...805..115V,2021A&A...647A..44V}.

One of the key issues related to thermal instability in gas and dust disks, in our opinion, is the question of its actual effect on disk morphology and the nature of accretion onto the star. The periodic nature of accretion, illustrated in the last section, arises when using the $\alpha$-parameterization of turbulent viscosity, which provides a positive feedback between the accretion rate and temperature. Meanwhile, the source of viscosity in protoplanetary disks has not yet been reliably established. Note that when using the $\beta$-parameterization of turbulent viscosity, where there is no dependence on temperature, no periodic accretion will occur within the viscous disk model,see~\cite{2001A&A...367.1087H}. Thus, the effect of thermal instability should be studied together with the question of the source of turbulent viscosity in the disk.

\section*{Funding}
The study was supported by the Russian Science Foundation grant No. 22-72-10029, https://rscf.ru/project/22-72-10029/.

\section*{ACKNOWLEDGMENTS}
The authors are grateful to the referee for valuable comments and suggestions for improving the article. We also express our gratitude to Yu.A. Fadeev, D.A. Semenov, L.I. Mashonkina, T.M. Sitnova for discussing the problem of calculating gas opacity.

\appendix
\section{Derivation of the formula for the midplane temperature of the circumstellar disk}

We consider a circumstellar disk in a state of thermal equilibrium. In the plane-parallel approximation, the thermal structure of such a disk in the vertical direction can be described by a system of moment transfer equations for thermal radiation:
\begin{align}
&\frac{dF}{dz} = c\rho\kappa_\text{P}(B-E)  \label{m1}\\
&\frac{c}{3} \frac{dE}{dz} = -\rho\kappa_\text{R} F \label{m2},
\end{align}
where $F$ is radiation flux, $E$ is density of radiation energy, $B = aT^4$ is radiation energy density at thermodynamic equilibrium, $a$ is radiation density constant, $T$ is medium temperature,
$c$ is speed of light, $z$ is vertical coordinate measured from the midplane, $\rho$ is medium density, $\kappa_\text{P}$ and $\kappa_\text{R}$ are Planck and Rosseland mean opacity coefficients. The equation~\eqref{m1} describes the change in flux due to the difference between the emission and absorption of radiation energy. The equation~\eqref{m2} relates the radiation flux to the energy density in the Eddington approximation.
The system of equations \eqref{m1}--\eqref{m2} is closed by the equation:
\begin{equation}
\frac{dF}{dz} = \rho S, \label{m3}
\end{equation}
according to which the thermal radiation flux is generated by some heating source $\rho S$, {where $S$ [erg s$^{-1}$ r$^{-1}]$ is defined as the heating power per unit mass.} Let us rewrite these equations using the surface density $\Sigma = \int\limits_{0}^{z}\rho(z^\prime)dz^\prime$ as a variable:
\begin{align}
&c\kappa_\text{P}(B-E) = S \label{n1} \\
&\frac{c}{3} \frac{dE}{d\Sigma} = -\kappa_\text{R} F \label{n2}\\
&\frac{dF}{d\Sigma} = S. \label{n3}
\end{align}

We will assume that the heating power $S$ originates from two processes: absorption of stellar radiation and viscous dissipation:
\begin{equation}
S=S_\text{star} + S_\text{vis}.
\label{S_tot}
\end{equation}
Heating power due to viscous dissipation of gas in the stationary approximation can be found as:
\begin{equation}
S_\text{vis} = \frac{\Gamma_\text{vis}}{\Sigma_{0}}
=\frac{3}{8\pi}\frac{GM\dot{M}}{R^3} \bigg/ \Sigma_{0},
\label{S_vis}
\end{equation}
where $M$ is the stellar mass, $\dot{M}$ is the accretion rate through the disk, $R$ is the distance from the star to the disk element in question,
$\Sigma_{0}$ is surface density from the midplane to the upper boundary of the disk, $G$ is gravitational constant. The use of~\eqref{S_vis} is also based on the assumption that the rate of viscous dissipation per unit volume is proportional to the density of the medium.
The heating of the disk by stellar radiation is found  using the formula:
\begin{equation}
S_\text{star} = \kappa_\text{F} F_0\,e^{-\dfrac{\kappa_\text{F}(\Sigma_{0}-\Sigma)}{\mu}},
\label{S_star}
\end{equation}
where $\kappa_\text{F}$ is the extinction coefficient averaged over the stellar spectrum, $F_0=\dfrac{L}{4\pi R^2}$ is the radiation flux from the star reaching the surface of the disk, $L$ is stellar luminosity, $\mu$ is
cosine of the angle between the direction to the star and the normal to the disk surface.
The formula \eqref{S_star} is derived from a formal solution of the radiative transfer equation under the assumption that the absorption coefficient $\kappa_\text{F}$ is constant along the vertical direction. In this case, we also neglect stellar radiation from the opposite surface of the disk. Accounting for disk heating by stellar radiation by introducing the $S_\text{star}$ function into the \eqref{n1}--\eqref{n3} system is based on the assumption that the disk radiates weakly in the visible range, i.e. this range weakly intersects with the range of thermal radiation of the disk itself.
We introduce the notation
\begin{equation}
\tau_\text{uv}=\dfrac{\kappa_\text{F}\Sigma_{0}}{\mu},
\end{equation}
which is the optical depth of the medium to stellar radiation up to the current position in the disk. Integration of the equation~\eqref{n3}, taking into account the expressions~\eqref{S_tot}--\eqref{S_star} and the condition that thermal radiation flux in the midplane is zero (due to the symmetry of the problem), gives:
\begin{equation}
F=\mu F_0\, e^{-\tau_\text{uv}}\left(e^{\dfrac{\tau_\text{uv}\Sigma}{\Sigma_{0}}}-1\right)
+S_\text{vis}\Sigma.
\label{nn3}
\end{equation}
In particular, on the disk surface, the thermal radiation flux is equal to:
\begin{equation}
F(\Sigma_0)=\mu F_0\, \left(1- e^{\tau_\text{uv}}\right) +S_\text{vis}\Sigma_0.
\label{nn4}
\end{equation}
By substituting the equation~\eqref{nn3} into the equation~\eqref{n2} and integrating the resulting equation from the midplane to the upper boundary of the disk, we can obtain a relation between the radiation energy density on the surface $E(\Sigma_0)$ and in the disk midplane $E(0)$:
\begin{equation}
E(\Sigma_0) - E(0) = - \dfrac{3\kappa_\text{R}\mu^2 F_0}{c\kappa_\text{F}} \left(1-\tau_\text{uv}e^{-\tau_\text{uv}} - \tau_\text{uv} \right)
-\dfrac{3\kappa_\text{R} S_\text{vis}}{2c} \Sigma_{0}^2.
\label{k2}
\end{equation}
When obtaining the relation~\eqref{k2}, it was assumed that $\kappa_\text{R}$ is constant along the vertical direction. As a boundary condition on the disk surface, we can use the relation:
\begin{equation}
F(\Sigma_0) =\eta\, c E(\Sigma_0),
\label{bound}
\end{equation}
where the coefficient $\eta$ depends on the assumed anisotropy of the outgoing thermal radiation. The value $\eta=0.5$ corresponds to isotropy over the upper hemisphere, while $\eta=1$ describes the case of strictly vertical radiation output. In what follows, we will set $\eta=0.5$. Combining the equations~\eqref{nn4}, \eqref{k2} and \eqref{bound}, one can obtain an expression for the radiant energy of thermal radiation in the midplane:
\begin{multline}
E(0) = \frac{F_0}{c}\left[
2\mu(1-e^{-\tau_\text{uv}}) + 3\mu^2 \frac{\kappa_\text{R}}{\kappa_\text{F}}
(1-e^{-\tau_\text{uv}}-\tau_\text{uv}e^{-\tau_\text{uv}})\right]+ \\
+\frac{2S_\text{vis}\Sigma_0}{c} \left(1+\frac{3}{4}\tau_\text{R} \right),
\label{ES0}
\end{multline}
where the Rosseland optical depth to thermal radiation is introduced:
\begin{equation}
\tau_\text{R} =\kappa_\text{R}\Sigma_0.
\end{equation}
The desired midplane temperature $T_\text{mid}$ is found from:
\begin{equation}
B(0)=a T_\text{mid}^4.
\label{midplanck}
\end{equation}
The value of $B(0)$, in turn, is expressed through the equation~\eqref{n1}, which, taking into account the value of the source function in the midplane~\eqref{S_tot}--\eqref{S_star}, takes the form:
\begin{equation}
B(0)= E(0) + \frac{\kappa_\text{F}F_0\,e^{-\tau_\text{uv}}}{c\kappa_\text{P}}
+\frac{S_\text{vis}}{c\kappa_\text{P}}.
\label{B0}
\end{equation}
Combining the equations \eqref{ES0}, \eqref{midplanck} and \eqref{B0}, we get:
\begin{multline}
a T_\text{mid}^4 =  \dfrac{\mu F_0}{c}\left[
2\left(1-e^{-\tau_\text{uv}}\right) + 3\mu \dfrac{\tau_\text{R}}{\tau_\text{uv}}
\left(1-e^{-\tau_\text{uv}}-\tau_\text{uv}e^{-\tau_\text{uv}}\right)\right]+  \\
+ \dfrac{\mu F_0\tau_\text{uv}}{\tau_\text{P}c} e^{-\tau_\text{uv}} +\dfrac{\Gamma_\text{vis}}{c}\left[\dfrac{
1+2\tau_\text{P}\left(1+\dfrac{3}{4}\tau_\text{R}\right)}{\tau_\text{P}}
\right],
\label{final}
\end{multline}
where the Planck mean optical depth with respect to thermal radiation is introduced:
\begin{equation}
\tau_\text{P} = \kappa_\text{P}\Sigma_0.
\end{equation}
The \eqref{final} equation can also be usefully expressed as follows:
\begin{equation}
\Lambda_\text{IR}=\Gamma_\text{star}+\Gamma_\text{vis},
\label{final2}
\end{equation}
{where $\Lambda_\text{IR}$ [erg c$^{-1}$ cm$^{-2}$] is cooling rate of midplane disk layers,
$\Gamma_\text{star}$ [erg c$^{-1}$ cm$^{-2}$] is heating rate by stellar radiation for the midplane layers, $\Gamma_\text{vis}$ [erg c$^{-1}$ cm$^{-2}$] is heating rate due to viscous dissipation:}
\begin{flalign}
&\Lambda_\text{IR}=
   \frac{4\tau_\text{P} \sigma T_\text{mid}^4}{1+2\tau_\text{P}\left(1+\dfrac{3}{4}\tau_\text{R}\right)}
   \label{lambda_ir} \\
&\Gamma_\text{star} =  \hspace{7.3cm}\nonumber\\
&=\frac{\mu F_0\tau_\text{P}\left[
   2\left(1-e^{-\tau_\text{uv}}\right) + 3\mu \dfrac{\tau_\text{R}}{\tau_\text{uv}}
   \left(1-e^{-\tau_\text{uv}}-\tau_\text{uv}e^{-\tau_\text{uv}}\right)
   +\frac{\tau_\text{uv}}{\tau_\text{P}} e^{-\tau_\text{uv}}
   \right]}{1+2\tau_\text{P}\left(1+\dfrac{3}{4}\tau_\text{R}\right)}
   \label{gamma_star} \\
&\Gamma_\text{vis}=\frac{3}{8\pi}\frac{GM\dot{M}}{R^3}.
   \label{gamma_vis}
\end{flalign}

Let us analyze the behavior of \eqref{final} in the absence of viscous heating when $S_\text{vis}=0$. For small optical depths with respect to stellar radiation $\tau_\text{uv}\ll1$ we get:
\begin{equation}
a T_\text{mid}^4 = \frac{\kappa_\text{F}}{\kappa_\text{P}}\frac{F_0}{c},
\label{lim1}
\end{equation}
where the temperature of the medium is determined by the ratio of the opacity of the medium to the stellar and thermal radiation. If the disk is optically thick to stellar radiation $\tau_\text{uv}\gg 1$ and the ratio $\kappa_\text{R}/\kappa_\text{F}$ is small (which is usually the case), then:
\begin{equation}
a T_\text{mid}^4 = \dfrac{2\mu F_0}{c},
\label{lim2}
\end{equation}
where midplane temperature depends only on the total flux of stellar radiation entering the disk. If the disk is optically thick to stellar radiation and there is viscous heating $S_\text{vis} \ne 0$, then:
\begin{equation}
a T_\text{mid}^4 = \dfrac{2\mu F_0}{c} +\frac{\Gamma_\text{vis}}{c}\left[\frac{
1+2\tau_\text{P}(1+\frac{3}{4}\tau_\text{R})}{\tau_\text{P}}
\right].
\label{lim3}
\end{equation}

\begin{figure}
\includegraphics[width=1\columnwidth]{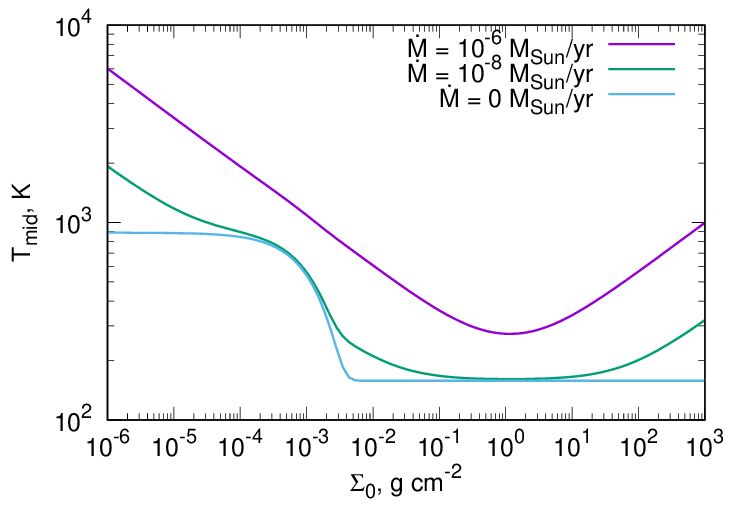}
\caption{Dependences of the midplane temperature on the disk surface density for different accretion rates $\dot{M}$ = 0, $10^{-8}$ and $10^{-6}$ $M_\odot$/year.}
\label{figA1}
\end{figure}

As an example, Fig.~\ref{figA1} shows the midplane temperature distributions obtained using the equation~\eqref{final}, depending on the surface density of the disk. The following parameters were used: $M=1M_{\odot}$, $L=1L_{\odot}$, $R=1$~au, $\mu=0.05$, $\kappa_\text{P}=\kappa_\text{R}=1$~cm$^2/g$, $\kappa_\text{F}=100$~cm$^2/g$.
These  dependences illustrate the limits obtained in \eqref{lim1} and \eqref{lim2}, and also show the effect of viscous heating.

\bibliographystyle{mn2e} 
\bibliography{maikbibl}

\end{document}